\documentclass[prb,showpacs,floatfix,amsfonts]{revtex4}
\usepackage{graphicx,graphics,color,epsfig}
\usepackage{bm}
\usepackage{amsmath}
\usepackage{amssymb}

\newcommand{\bR}{{\bf R}}

\newcommand{\bk}{{\bf k}}

\newcommand{\br}{{\bf r}}

\newcommand{\beqa}{\begin{eqnarray}}
\newcommand{\eeqa}{\end{eqnarray}}

\begin{document}
\preprint{}
\title{Local Strong Coupling Pairing in  $D$-Wave Superconductor with  Inhomogeneous  Bosonic Modes
}
\author{A. V. Balatsky and  Jian-Xin Zhu}
\affiliation{Theoretical Division, Los Alamos National Laboratory,
Los Alamos, New Mexico 87545}

\date{March 7 2006}

\begin{abstract}
Recent local tunneling data indicate strong nanoscale inhomogeneity
of superconducting gap in high temperature superconductors. Strong
local  nanoscale inhomogeneity in the bosonic scattering mode has
also been observed in the same samples. We argue that these two
inhomogeneities directly related to each other. To address local
boson scattering effects,  we develop a local strong coupling model
of superconducting pairing in a coarse grained superconducting
state. Each patch is characterized by local coupling to the bosonic
mode as well as by local mode energy. We find that local gap value
on each patch grows with the local strength of electron-boson
interaction.  At the same time local gap value decreases with the
local boson mode energy, an observation consistent with the
tunneling experiments.  We argue that features  in the tunneling
spectrum due to boson scattering are consistent with experimentally
observed spectra. We also address the $^{16}O$ to $^{18}O$ isotope
substitution. Since both coupling constant and boson energy could
change upon isotope substitution, we prove that interplay between
these two effects can produce results  that are very different from
conventional BCS model.
\end{abstract}
 \maketitle

\section{Introduction}

Electron-boson interaction is at the center of the pairing
interaction in conventional superconductors. Pioneering work of
Eliashberg,~\cite{Eliashberg:60} McMillan and
Rowell,~\cite{McMillan:65}, and Scalapino~\cite{Scalapino:69}
conclusively proved that the non-BCS features in the tunneling
spectra in conventional superconductors are directly related to the
electron-boson coupling and ultimately to the formation of the
superconducting state. Experimental evidence of the
``strong-coupling'' features~\cite{Carbotte:90} in the tunneling
spectra was clearly connected to the known phonon spectra in these
materials.~\cite{Giaever:60} On the other hand the number of
superconducting materials where electron-boson coupling has been
seen was small and initial weak coupling approach of
Bardeen-Copper-Schrieffer theory was successful in predicting all
the measured properties in these materials. The few superconductor,
like Pb and Sn, with clear electron-phonon features in the spectrum
were called ``bad actors'' because of deviations from BCS
predictions.

In this paper we will focus on high temperature superconductors
(high-$T_c$). In high-$T_c$ materials  situation is very different.
Tunneling spectra in all of these materials clearly deviates from
the mean field  BCS $d$-wave tunneling density of states (DOS), see
e.g. Fig.(\ref{FIG:dos}). In this sense all of the high-$T_c$
materials are ``bad actors''.

There has been substantial evidence for strong quasiparticle
renormalization from
tunneling~\cite{Huang89,Mandrus91,Renner95,Renner96,DeWilde98,
Yurgens99,Zasadzinski00,Zasadzinski01} and  angle-resolved
photoemission
(ARPES)~\cite{Dessau91,Ding96,Shen93,Campuzano99,Bogdanov00,Kaminski01,
Lanzara01,Johnson01,Zhou03,Kim03,Gromko03,Sato03,Cuk04,Zhou05}
experiments. The strong quasiparticle renormalization has been
suggested as a manifestation of strong electronic coupling to
collective
modes.~\cite{Dahm96,Shen97,Norman97,Norman98,Abanov99,Eschrig00,Norman01,
Manske01,Abanov03,Sandvik04,Devereaux04,Bogdanov00,Kaminski01,
Lanzara01,Johnson01,Zhou03,Kim03,Gromko03,Sato03,Cuk04,Zhou05}
Observed strong electron-electron correlations  are clearly
important for mechanism of superconductivity. At the same time the
physics of electron-boson coupling in these materials is crucial for
understanding of the pairing in these materials. We would argue that
these features of the tunneling spectra are due to strong
electron-boson coupling.

Recent scanning tunneling microscope (STM) experiments reveal strong
electronic inhomogeneity.~\cite{Cren00,
Howard01,Pan01,Lang02,Fang04,McElroy05} More recent  experiments
with inelastic electron tunneling spectroscopy (IETS STM) have
allowed one to directly observe inelastic tunneling features in
Bi$_{2}$Sr$_{2}$CaCu$_{2}$O$_{8+\delta}$ (BSCCO) high-$T_c$
materials. These experiments  have shown that bosonic modes that
produce strong coupling features in tunneling spectra are also
inhomogeneous on the scale of $20-50 \AA$.\cite{Jinho1}  Nontrivial
correlations between local gaps and local boson mode energies were
observed.

Here we follow the notion that deviations from the BCS tunneling DOS
in these materials are caused by the strong coupling effects due to
electron-boson interaction in these materials. We present a strong
coupling theory of $d$-wave superconductor where superconducting
state arises as a result of a pairing mediated by bosonic modes,
that is attractive in $d$-wave channel.

In this regard, the treatment is analogous to the conventional
Eliashberg-McMillan-Rowell approach. The main departure from the
conventional approach is that we explicitly allow inhomogeneity in
the electron-boson coupling strength and the bosonic mode energy.
Typical size of inhomogeneity we will assume is on the order of
$20-50\AA$. We will assume that the pairing is local and determined
by local values of coupling strength and mode energy at the given
patch. This approximation allows us to simplify the calculation
dramatically. If one takes guidance from the data,  it is clear that
inelastic tunneling features and superconducting gaps are rather
local, and there is no ``self-averaging''  seen in the tunneling
spectra. In other words, local approximation might be a good
starting point for this kind of analysis.

We solve self-consistently the Eliashberg equations on each patch
and find local $d$-wave order parameter. Random Gaussian
distributions of the local coupling constant and local mode energy
are considered. As the result, local order parameter (OP) values are
random maps that  correlate with the input parameters. We find that
the local OP positively correlate with the coupling constants. One
of the most important findings of this local formalism is that we
find indeed anticorrelation between the typical local boson
frequency and the superconducting OP.
  This negative correlation  is a direct consequence of the strong
  local
coupling nature of the pairing we assumed. The IETS-STM experiment
by Lee {\em et al.}~\cite{Jinho1} has indeed shown a negative
correlation between gaps and IETS mode energy.

Using our local model we  also address   isotope effect. The most
commonly used isotope substitution for high-$T_c$ materials is
$^{16}O$ by $^{18}O$  substitution. Isotope effect was studied in
the past by measurements other than STM tunneling. Here we point to
the important papers in this regard by Shen and Lanzara groups
~\cite{Lanzara01}  that argue how isotope effect can change
 both the characteristic bosonic frequency and the coupling
 constant to the local bosonic mode.\cite{Bogdanov00,Lanzara01,Zhou03,Cuk04,Zhou05}
 We find that isotope shift in these two quantities
 can either work together to mutually enhance the
 superconducting gap, or they can work against each other partially cancelling
  and therefore making net  isotope effect small. Hence we conclude that
  the naive arguments  about
 the ``smallness''  of isotope effect with substitution of  $^{16}O$ by
 $^{18}O$ are  misleading.

While in this paper we focus on electron-boson interaction, in real
systems this interaction {\em contributes} to the pairing in these
materials. We would like to make it clear that we believe pairing in
high-$T_c$ materials is a result of interplay between strong
electron-electron correlations \cite{Nunner05} and electron-boson
interaction. To address the effects of spatial inhomogeneity of
tunneling IETS spectra we focus only on electron-boson coupling,
ignoring electron-electron interaction part that will not produce
IETS features. It is not clear how high the transition temperature
would be assuming only electron-boson pairing. We leave this
question for a separate investigation.

The outline of the paper is as follows: In Sec.~II, we will
introduce local strong coupling formalism, and  outline the details
of the formalism starting with the general inhomogeneous pairing
theory in reals space. In Sec.~III, we will present results from
solving self-consistently the strong coupling equations. In Sec.~IV
we will discuss the isotope effect for inhomogeneous superconductor.
We conclude in Sec.~V.

\section{Local pairing  formalism}

We start with the strong coupling analysis in the local pairing
limit. Locality here would be understood in the sense of a coarse
grain approach where we assume that the typical sizes of grains are
on the order of the coherence length of superconductor $\xi \sim 20
\AA$. This assumption is consistent with the STM observed
granularity in Bi2212~. \cite{Jinho1,McElroy05}

We will present a detailed description of local strong coupling
theory with the steps in the logic  outlined. Some of the points are
well known  but we keep them in for completeness. The formulas are
very similar to the standard Eliashberg discussion except we want to
stay with the real space description.
~\cite{McMillan:65,Eliashberg:60,Scalapino:69}

Taking the STM data as guidance we can imagine that we take a coarse
grained vies of the sample. We are taking the given field of view
and pixelizing it in a set of boxes with characteristic size of $20
\times 20 \AA$. Each of the pixels will be assigned  three
variables: order parameter $W(\bR)$, bosonic mode energy
$\Omega_0(\bR)$  and local coupling constant $g(\bR)$. Coupling
constant and bosonic mode energy variables are randomly drawn from
the given distribution. We calculate the local order parameter
self-consistently.

To start we  write down the Green's function equations in the Nambu
space, assuming no translational invariance. We consider the case of
both electron-electron and electron-boson interactions being
present. The electron-boson coupling term in the Hamiltonian
assuming no translational invariance is: \beqa H_{e-lattice} = \int
d\br d\br' d\br'' c^{\dag}_{\br \sigma} c_{\br'
\sigma}(b^{\dag}_{\br''} + b_{\br''}) g(\br, \br', \br'') \;,
\label{EQ:eb1} \eeqa
 where $g(\br, \br', \br'')$ is the coupling constant and $x(\br) = b_\br + b^\dag_\br$ is the boson field.
If we assume that only electronic density coupled to the lattice
degrees of freedom, then coupling constant $g(\br, \br', \br'')
 = g(\br) \delta_{\br, \br'}
\delta_{\br, \br''}$. The bosons that couple to the electronic
density on the other hand are taken to be on the bonds connecting
nearest neighbor sites, or alternatively to reside on the dual
lattice. We will focus on the local bosonic mode coupling. This seem
to be a general enough situation that we believe will capture the
relevant physics to be  addressed.

Next we consider the second order terms in the effective action that
would look like \beqa \int d{\br \br' \br''} d{\br_1 \br_1''
\br_1'''} \int _{0}^{\beta} d \tau d\tau_1g(\br,\br', \br'')
g(\br_1, \br_1', \br_1'') c^{\dag}_{\br \sigma}(\tau) c_{\br'
\sigma} (\tau)c^{\dag}_{\br_1 \sigma_1}(\tau_1) c_{\br_1'
\sigma_1}(\tau_1) x_{\mathbf{r^{\prime\prime}}}(\tau)
x_{\mathbf{r_{1}^{\prime\prime}}}(\tau_{1}) \;,
 \label{EQ:action1} \eeqa
 here we assume that phonon propagator $B(\br -\br_1,\tau - \tau_1) =
 -
 \langle T_{\tau} x(\br) (\tau) x(\br_1)(\tau_1)\rangle$ is
{\em local} given the STM data that indicate the strong local
variations of the bosonic excitations as seen in IETS STM
\cite{Jinho1}. Thus we assume \beqa -\langle T_{\tau} x(\br)
x(\br_1)\rangle = B(\br,\tau) \delta_{\br, \br'} \label{EQ:eb2}
\eeqa

Then in the self-energy terms electron-boson interaction would
produce the term that we would  look as $ g(\br)g(\br''')
B_{\Omega_m}(\br, \br'|\br'', \br''')$.  We again stress here that
both coupling constant $g$ and boson energy will be assumed
inhomogeneous. This situation is qualitatively different from weak
coupling approach where only single effective coupling constant
 will be position dependent.

Equation for Green's functions  in Matsubara frequency $\omega_n =
\pi(2n+1)k_BT$ becomes: \beqa \left(
  \begin{array}{cc}
    {\hat \varepsilon}_\br - S(\br,\br',i\omega_{n}) & W(\br,\br',i\omega_n) \\
    W^*(\br,\br',i\omega_n) & -{\hat \varepsilon}_\br - S(\br, \br',i\omega_n) \\
  \end{array}
\right) \bigotimes {\hat G}(\br',\br'',i\omega_{n}) = \mathbf{1}
\delta(\br - \br'') \label{EQ:matrix1} \eeqa with $\bigotimes$
understood as a convolution in real space. ${\hat \varepsilon}_\br$
is the kinetic energy operator, that in momentum space will have the
form $\varepsilon(\bk)  = -t (\cos(k_x a) + \cos(k_y a)) - \mu$, $
\mu$ being the chemical potential. $\mathbf{1}$ being the indentity
matrix,and $\hat{G}(\br, \br',i\omega_n)$ being a matrix in Nambu
space with relevant components $\hat{G}(\br, \br',i\omega_n)_{11} =
G(\br,\br',i\omega_n)$ and $\hat{G}(\br, \br',i\omega_n)_{12} =
F(\br,\br',i\omega_n)$ being the normal and anomalous Green's
functions.

We also explicitly keep the normal self energy in the Gorkov
equations: $S(\br, \br',i\omega_n)$ that renormalizes the normal
selfenergy propagators.

Normal and anomalous selfenergies are defined selfconsistently
through:

\beqa S(\br, \br', i\omega_n) = -T \sum_m \int d \br'' d\br
'''V_{\omega_n - \omega_m}(\br, \br'|\br'', \br''') G(\br'',\br''',
i\omega_m) \label{EQ:S1} \eeqa

\beqa W(\br, \br', i\omega_n) = -T \sum _m \int d\br'' d\br'''
V_{\omega_n - \omega_m}(\br,\br'|\br'',\br''') F(\br,\br',i\omega_n)
\label{EQ:F1} \eeqa

where the effective pairing interaction
$V_{\Omega_m}(\br,\br'|\br'', \br''')$ is given by the combination
of both direct and lattice-vibration-induced electron-electron
interaction:
 \beqa
V(\br,\br'|\br'',\br''')_{\Omega_n} = V^{ee}(\br,\br'|\br'',\br''')
- g(\br)g(\br''') B_{\Omega_m}(\br,\br'|\br'',\br''') \label{EQ:V1}
\eeqa $V$ is defined on Matsubara bosonic frequency $\Omega_m = 2\pi
m k_BT$, assuming local coupling $g$. Here $V^{ee}$ is an
electron-electron interaction written in real space (below we will
assume that this term might be inhomogeneous as well). We assume
$V^{ee}$ to be weakly frequency dependent.

Next we  introduce  local (on a coarse grained scale)  description
for the properties of superconductor. For any discussion of the
local nature of pairing we will need to keep track of the relative
coordinate and center of mass coordinates. For example, consider the
pairing amplitude $W(\br,\br',\omega_n)$ that describes the pairing
amplitude of two particles at sites$\br$ and $\br'$. It is
convenient to introduce the center of mass and relative coordinates
for the pairing field and for the kernel $g^2B$ \beqa \bR = 1/2(\br
+ \br'), \bR_1 = 1/2(\br_1 + \br_1')
\nonumber\\
\tilde{{\br}} = \br-\br', \tilde{\br_1} = \br_1 - \br_1'
\label{EQ:coordinates1} \eeqa

for simplicity of notation we will drop ~  sign  in $\tilde{{\br}}$
hereafter with understanding that capital coordinate label would
mean center of mass coordinates and small coordinate label would
mean the relative coordinates. In thee coordinates
$W(\br,\br',\omega)$ becomes $W(\bR, \br,\omega)$. And similar
expressions for $S, \Delta$ etc. The interaction term we will  also
have a factorizable form

\beqa V_{\Omega_m}(\br, \br'|\br_1, \br_1') = V_{\Omega_n}(\bR,
\br|\bR_1,\br_1) \label{EQ:Veff1} \eeqa

 In the case of homogeneous
pairing $W$ is independent of $\bR$. We will focus on the
inhomogeneity of $W(\bR,\br,\omega)$ as a function of $\bR$.

We introduce the basis functions for d-wave channel and  ignore any
other pairing channels. This is not a principal assumption but a
useful one that allows us to greatly simplify  equations.

We have therefore  \beqa W(\bR,\br,\omega) & = & W(\bR,\omega)\eta(\br) \nonumber\\
 V_{\Omega_n}(\bR, \br|\bR_1,\br_1) & = & V_{\Omega_n}(\bR,\bR_1)
 \eta(\br)\eta(\br_1)
 \label{EQ:separable1}
 \eeqa

 and similar for $S(\br, \br', \omega_n)$ etc.

Here $\eta(\br)$ is a real space representation of the basis
function that has a d-wave character. The simplest way to present is
is to take a function that is nonzero at nearest neighbors of
site$\br$ that has a d-wave signature: $\eta(\br) = \sum_{\delta =
\pm x, \pm y} (-1)^\delta \delta_{\br, \br+\delta}$ on the lattice.
In the continuum we would have to deal with the gradient operator:
$\eta(\br) \sim (\partial^2_x - \partial^2_y)$. In the momentum
space it will be a simple $\eta(\bk) = \cos(k_x a) - \cos(k_y a)$.

It is also convenient to introduce mixed representation where we use
Fourier transform  for the relative coordinate. Then

\beqa W(\omega_n,\bR, \bk)  & = &W(\omega_n,\bR)\eta(\bk)
\nonumber\\
S(\omega_n, \bR,\bk) & = & S(\omega_n,\bR){\bf 1}_{\bk} \nonumber\\
V_{\Omega_m}(\bR, \bk|\bR_1, \bk_1)  & = & V_{\Omega_n}(\bR,\bR_1)
\eta(\bk) \eta(\bk_1) \label{EQ:FT1} \eeqa

We consider d-wave channel only in assuming a simple factorizable
approximation. This is definitely an oversimplification since the
inhomomgeneous system would admit the mixture between components.
One can always include the mixing in a more detailed approach. In
practice this mixture could be modest. For us the main focus here
would be on the real space modulations of the gap function $W(\bR,
\omega_n)$, electron-lattice and electron-electron interactions
$V_{\Omega_m}(\bR, \bR')$. We proceed with this simplifying
assumption that would make our discussion more transparent.

We can rewrite the selfenergies $S,W$ Eq.(\ref{EQ:S1},\ref{EQ:F1})
upon projection on the d-wave channel as:

\beqa W(\bR, \omega_n) = -T\sum_m \int d\bR' V_{\omega_n -
\omega_m}(\bR, \bR') \langle F(\omega_m, \bR', \bk')\rangle_d
\label{EQ:F2} \eeqa
 and similarly for $S$:

 \beqa
S(\bR, \omega_n) = -T\sum_m \int d\bR' V_{\omega_n - \omega_m}(\bR,
\bR') \langle G(\omega_m, \bR', \bk')\rangle \label{EQ:S2} \eeqa

 with
 \beqa \langle F(\omega_m, \bR', \bk')\rangle_d =
 \int d\bk' \eta(\bk')F(\omega_m, \bR', \bk')
 \label{EQ:F3}
 \eeqa

  \beqa
\langle S(\omega_m, \bR', \bk')\rangle = \int d\bk' S(\omega_m,
\bR', \bk') \label{EQ:S3} \eeqa

We  focus on the gap equation hereafter. We find that normal
self-energy $S(\omega_n, \bR,\bk)$, at most leading to mass
renormalizations on the scale unity.  The fermi surface average
correction due to normal self energy corrections  is small and hence
we ignore it. This allows us to keep only  d-wave projected part of
the interaction in Eq.(\ref{EQ:FT1}).

From the solutions of the Green's function we have self consistently
defined $F, G$:

\beqa F(\omega_n, \bR, \bk) = \frac{W_{\omega_n}(\bR)
\eta(\bk)}{(i\omega_n  - S(\omega_n, \bR))^2 - \varepsilon(\bk) -
W^2(\omega_n,\bR)\eta^2(\bk)} \label{EQ:F3} \eeqa

\beqa G(\omega_n,\bR, \bk) = \frac{i\omega_n -
S(\omega_n,\bR)}{(i\omega_n - S(\omega_n, \bR))^2 - \varepsilon(\bk)
- W^2(\omega_n,\bR)\eta^2(\bk)} \label{EQ:S3} \eeqa

Gap in the quasiparticle spectrum is determined as \beqa
\Delta(\omega_n, \bR) = \frac{W(\omega_n,\bR)}{1 -
S(\omega_n,\bR)/i\omega_n} \label{EQ:Gap1} \eeqa

These equations are written in general form. We take  $S = 0$ below.

Equations
Eq.(\ref{EQ:F2},\ref{EQ:S2},\ref{EQ:F3},\ref{EQ:S3},\ref{EQ:Gap1})
are the main result of this section. These equations are quite
general and describe the inhomogeneous superconducting state in the
presence of  inhomogeneous pairing interaction.

  These equations are
similar to the Eliashberg equations considered for a homogeneous
superconductor. Here we focus on the spatial dependence of the
superconducting properties like gap in the spectrum and pairing
interaction $V_{\Omega_m}(\bR,\bR')$.

\subsection{Local Approximation}

We can make further progress if we will make some additional
assumptions. We will assume that the kernel in
Eq(\ref{EQ:F2},\ref{EQ:S2}) {\em local}. Again,  this locality
should be understood in a coarse graining sense with typical length
scale for
 coarse graining to be on the order of superconducting coherence length $\xi \sim 20 \AA$.
This length scale is compatible with the observed inhomogeneities in
the tunneling gap and bosonic frequency, as imaged with
STM.\cite{Jinho1}

  Local on-site pairing kernel would be incompatible with the d-wave
character of the pairing we assumed here. For a moment we will focus
on the electron-lattice part of the kernel. It contains an effective
coupling $g$ and boson propagator $B$ as a single combination we
label $g^2B$. We will assume  local approximation for {\em both
couling constant $g$ and bosonic frequency}.  Thus:
 \beqa V_{\Omega_m}(\bR,\bR') = V^{ee}(\bR)\delta_{\bR \bR'}  -
 g(\bR)g(\bR')B_{\Omega_m}(\bR)\delta_{\bR \bR'}
 \label{EQ:local3}
 \eeqa

Where, following standard Eliashberg
 approach \cite{Eliashberg:60,McMillan:65,Scalapino:69},  we will
assume that the bosonic spectral density (on the real frequency
axis) would have a local character:
 \beqa Im(g^2B(\bR,\bR',\Omega))
= \pi g^2(\bR) [\delta(\Omega - \Omega_0(\bR)) -  \delta(\Omega +
\Omega_0(\bR))]\delta_{\bR \bR'}\label{EQ:local4} \eeqa

\begin{figure}[th]
\centerline{\psfig{file=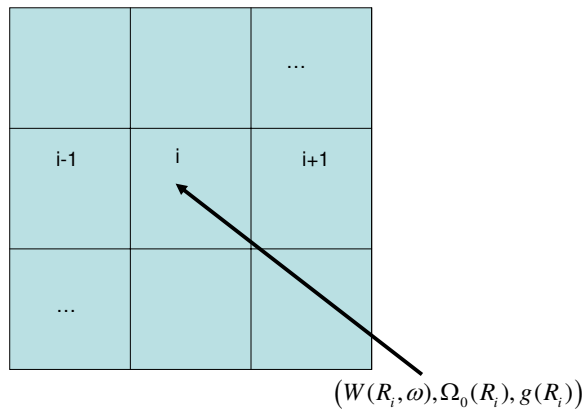,width=8cm}} \caption{
Illustration of our coarse graining approach to the local strong
coupling solution of Eliashberg equations. We pixelize a finite
system with the size of the patch to be on the order of $\xi$. Then
we solve the selfconsistent strong pairing equations on each patch.
The outcome of the solution is the set of local value for d-wave
order parameter, that depends on the local value of boson frequency
and local coupling constant. We present these values as a pixel
dependent vector with these three components $(W(\bR_i,\omega),
\Omega_0(\bR_i), g(\bR_i))$. } \label{FIG:Pixel1}
\end{figure}

  With the recent STM experiments we now
have an independent experimental measure of the local bosonic energy
$\Omega_0(\bR)$ as a function of position and doping \cite{Jinho1}.
Typically, these energies are  randomly distributed in a sample with
characteristic variations on the length scale on the order of $20 -
50 \AA$ and thus are consistent with our assumption of locality. The
sample averaged frequency $\Omega_0 = \langle \Omega_0(\bR) \rangle
$ is essentially doping independent and is about 52 meV, while
distribution ranges between 40-70 meV {\em for the observed bosonic
modes in STM experiments},  see Fig(\ref{FIG:Pixel1}).  We do not
have a similar experimental information on coupling constant.

In practice, of course, only  the total kernel $V$ will enter into
the self-consistency equations and we would not be able to
differentiate the effects of inhomogeneity in the  $W$ due to
electron-electron versus electron-lattice interaction
inhomomgeneity. There is  one important distinction however between
electron-electron vs electron-lattice coupling. Electron-electron
interaction being essentially frequency independent can not produce
features outside the coherence peaks. Electron-lattice coupling on
the other hand will produce the features  in local tunneling
characteristics at $\omega(\bR) = \Delta(\bR) + \Omega_0(\bR)$.
Hence the  local tunneling characteristics would allow us to extract
the inhomogeneous values of the bosonic modes, at least in
principle. In practice one would have to deal with rather large
signal- noise uncertainties but the local bosonic energy extraction
from the dI/dV data can be done \cite{Jinho1}.

The local version of equations Eq(\ref{EQ:F2},\ref{EQ:S2}) now would
take a form:

\beqa W(\bR, \omega_n) = -T\sum_m  V_{\omega_n - \omega_m}(\bR)
\langle F(\omega_m, \bR, \bk')\rangle_d \label{EQ:F4} \eeqa
 and similarly for $S$:

 \beqa
S(\bR, \omega_n) = -T\sum_m V_{\omega_n - \omega_m}(\bR) \langle
G(\omega_m, \bR, \bk')\rangle \label{EQ:S4} \eeqa

and one can recognize standard Eliashberg equations that are now
written locally, patch by patch.  Hereafter we will approximate
$S(\bR, \omega_n) = (1 - Z_{\omega_n}) i\omega_n$, i.e. local Z
factor normalization only. In practice we know that effective mass
renormalization in high-Tc materials is not very large and at most
$Z \sim 2$,  hence  the effects of the normal self energy on
quasipatrticle dispersion  would be minor. It is the pairing
interaction contribution from electron-boson interaction that we
will be paying attention to. These equations in the homogeneous case
were well analyzed. \cite{Eliashberg:60,McMillan:65,Scalapino:69}

\subsection{Weak Coupling Approximation}

Now we will consider the weak coupling limit of these equations,
namely the case when the pairing amplitude $W$ and normal
self-energy corrections $S$ are small compared to the typical
bosonic energy, $\Omega_0 = \langle \Omega(\bR)\rangle$.  The most
natural region to make the weak coupling approximation would be in
the overdoped regime where both superconducting gap $\Delta$ and
$T_c$ decrease with increased doping. There are no competing orders
in the overdoped regime
 that would make the analysis more
complicated. Competition with  the other orders, like charge ordered
state and pseudogap would make our analysis in terms of a single
superconducting gap inaccurate. Therefore the analysis presented
below assumes that  we are dealing with optimally doped to overdoped
samples.

 Summation over Matsubara frequency in Eq.(\ref{EQ:F4},\ref{EQ:S4}) is treated in a standard way by
using $T \sum_n f(\omega_n) = \oint dz \tanh(\frac{\beta z}{2})
f(z), \beta = 1/k_BT$. This integral is reduced to the integral over
spectral density for the effective interaction that we will assume
to be local:

\beqa V(\Omega) & = & \int_{-\infty}^{\infty} d x \frac{ImV(x)}{x-\Omega} \nonumber\\
Im(g^2_{\Omega}(\bR,\bR')B_{\Omega}(\bR,\bR')) & = & g^2(\bR)
(\delta(\Omega - \Omega_0(\bR)) - \delta(\Omega + \Omega_0(\bR))
\delta_{\bR, \bR'} \nonumber\\
 Im V^{ee}(\bR,\bR',\Omega) & = & V^{ee}(\bR) \delta_{\bR, \bR'} , \ \ |\Omega| \leq \Omega_c
 \nonumber\\
 Im V_{\Omega} (\bR)& = & Im V^{ee}(\bR) - Im g^2B_{\Omega}(\bR)
 \label{EQ:local5} \eeqa

where we have parametrized the position dependent electron-lattice
interaction by local coupling constant $g^2$ and local boson
frequency $\Omega_0(\bR)$. Electron-electron interaction is assumed
to be frequency independent up to cut-off frequency $\Omega_c \gg
\Omega_0$.

Analysis we will implement here is essentially the same as the one
used in standard Eliashberg approach. We use the  d-wave projected
propagator $F$ and integrate over the quasiparticle energies
$\varepsilon(\bk)$ to find from Eq(\ref{EQ:F4},\ref{EQ:S4}):

\beqa W(\bR,\Omega) & = & - N_0 \int dx  \int_0^{\infty} d \Omega '
\tanh(\frac{\beta \Omega }{2}) \frac{Im (V_x(\bR))}{x-\Omega +
\Omega'} Re (\langle\frac{W(\bR,
\Omega')}{\sqrt{(Z_{\Omega'}\Omega')^2 -
W^2(\bR, \bk, \Omega')}}\rangle_{\bk}) \nonumber\\
& = & \int_0^{\infty} d \Omega ' N_0 \frac{2 g^2(\bR)(\Omega_0(\bR)
+\Omega')}{(\Omega_0(\bR)+\Omega')^2 -(\Omega)^2 + i\delta} Re
(\langle\frac{W(\bR, \Omega')}{\sqrt{(Z_{\Omega'}\Omega')^2 -
W^2(\bR, \bk, \Omega')}}\rangle_{\bk}) \label{EQ:local6}
 \eeqa
In the weak coupling limit for small coupling  $g^2B(\omega, \bR)$
we can develop a {\em local BCS pairing approximation for this local
strongly coupled superconductor} by approximating $W(\omega,\bR) =
W(\omega = 0, \bR) \Theta(\Omega_0(\bR) - \omega), \omega > 0$ and
similar step like cut off at negative $\omega$.

 We will focus on the low energy
part of $W(\bR, \Omega)$ that will contain real part of the gap
only. Then a)  the integral over $x$ is trivial since we assumed
spectral function $Im V$  to be a delta function in frequency. We
assume $T\rightarrow 0$ and hence $\tanh(\frac{\beta \Omega }{2}) =
1$. b) we assume that electron-electron part will produce
normalizations on the normal channel and also will produce a d-wave
pairing. Then  we  take a low frequency limit of this equation
$\Omega \rightarrow 0$ and limit the integral over $\Omega'$ over
the range $\Omega' \leq \Omega_0$ since the kernel is attractive
only in this range.   The weak coupling limit therefore would read
as:

\beqa W(\bR,\Omega = 0) & =& \int_0^{\Omega_0(\bR)} d \Omega ' N_0
\frac{2 g^2(\bR)(\Omega_0(\bR) + \Omega')}{(\Omega_0(\bR)+\Omega')^2
-(\Omega)^2 + i\delta}\nonumber\\
 Re (\langle\frac{W(\bR,
\Omega=0)}{\sqrt{(Z_{\Omega'}\Omega')^2 - W^2(\bR, \bk, \Omega
=0)}}\rangle_{\bk}) \label{EQ:local7} \eeqa

and ultimately we obtain the {\em local version} of BCS equation for
$W(\bR) = W(\bR,\Omega = 0)$:

\beqa W(\bR) =  \int_0^{\Omega_0(\bR)}d \Omega ' N_0 g_{eff}(\bR) Re
(\langle\frac{W(\bR)}{\sqrt{(Z_{\Omega'}\Omega')^2 - W^2(\bR, \bk,
\Omega =0)}}\rangle_{\bk}) \label{EQ:local8} \eeqa

with \beqa g_{eff}(\bR) = 2 \int d \omega \frac{g^2 (\omega,
\bR)B(\omega, \bR)}{\omega} \label{EQ:geff1} \eeqa
 is a weak coupling limit coupling constant in Eliashberg theory.

 Using the local approximation for
 the spectral density Eq.(\ref{EQ:local5}), we have

 \beqa
 g_{eff} = 2 \frac{g^2(\bR)}{\Omega_0(\bR)}
 \label{EQ:geff2}
 \eeqa

   Here we explicitly assume that $g(\bR)$ and $\Omega(\bR)$ are {\em
 independent} random distributions. This assumption is natural if we
 allow these two quantities to be set by local environment in the
 crystal and we do not assume here that coupling constant $g(\bR) \sim
 (M \Omega(\bR))^{1/2}$ as would be the case for quantized extended collective
 modes.  We thus arrive at
the local gap equation for $\Delta(\bR) = W(\bR)/Z(\bR)$:

\beqa \Delta(\bR) = \Omega_0(\bR) \exp(-\frac{1}{N_0 g_{eff}(\bR)})
\label{EQ:local9} \eeqa

Equation Eq.(\ref{EQ:local9}) is applicable in the weak coupling
limit and therefore can be viewed only as a qualitative result. For
any distribution of the coupling constants for small enough average
value $<g^2B>$ there will be regions where this coupling constant is
not smaller than $1$ and hence weak coupling analysis would fail in
those regions. Nevertheless it is useful in that it allows us to
analyze the results of numerical calculations, see below and compare
numerical results with the locally observed quantities. With this
caveat in mind we will consider the implications of the
Eq.(\ref{EQ:local9}) for our analysis of the local pairing.

We  find immediately three important consequences of the local
pairing approximation: 1) Effective coupling constant is a function
of the local boson mode energy. Effective coupling constant
$g_{eff}(\bR)$ is inversely proportional to the mode energy
$\Omega_0(\bR)$. This result  implies that there is a {\em  direct
negative  correlation} between local gap and local bosonic mode
energy.  Similar direct negative correlation is observed in the STM
experiments on inelastic tunneling spectroscopy.\cite{Jinho1} 2)
Another implication of this result is that the isotope efffect will
affect both the prefactor and coupling constant in
Eq.(\ref{EQ:local9}). This point will be discussed more below.
3)Finally, from this simple equation we can find local effective
coupling constant $N_0g_{eff}(\bR)$ as:

\beqa N_0g_{eff}(\bR) = -
\frac{1}{\ln(\frac{\Delta(\bR)}{\Omega_0(\bR)})} \label{EQ:local10}
\eeqa

This equation contains two experimentally observable quantities:
$\Delta(\bR)$ and $\Omega_0(\bR)$. We therefore can build the real
space map of effective coupling constant.



\section{Numerical Results and Discussions}

Here  we will discuss   the numerical solutions of local Eliashberg
equation. From Eqs.(21) and (22) together with Eqs.(16) and (17), it
can be written explicitly as:
\begin{subequations}
\begin{eqnarray}
i\omega_{n} S(\mathbf{R},i\omega_n) & = & - \frac{1}{N_{L}\beta}
\sum_{\bf{q}} \sum_{i\Omega_{m}} \frac{M_{eff}(i\Omega_{m},\bR)
}{\Pi(\mathbf{R},\mathbf{q};i\omega_{n}-i\Omega_{m})}
[(i\omega_{n}-i\Omega_{m})Z(\mathbf{R},\mathbf{q};i\omega_{n}-i\Omega_{m})+\xi_{\mathbf{q}}\;,
\nonumber\\
W(\bR,i\omega_n) &=& - \frac{2}{N_{L}\beta} \sum_{\mathbf{q}}
\sum_{i\Omega_{m}} \frac{M_{eff}(i\Omega_{m},\mathbf{R})
}{\Pi(\mathbf{R},\mathbf{q};i\omega_{n}-i\Omega_{m})}
W(\mathbf{R},\mathbf{q};i\omega_{n}-i\Omega_{m})\;,
\end{eqnarray}
\end{subequations}
where
\begin{equation}
M_{eff}(\mathbf{R},\mathbf{k};i\Omega_{m}) = [V_{ee}(\mathbf{R}) -
g^{2}(\mathbf{R}) B(\mathbf{R},i\Omega_{m})] \eta_{\mathbf{k}}\;
\end{equation}
with
\begin{equation}
B(\mathbf{R};i\Omega_{m}) =
\frac{1}{i\Omega_{m}-\Omega_{0}(\mathbf{R})}
-\frac{1}{i\Omega_{m}+\Omega_{0}(\mathbf{R})} \;,
\end{equation}
and
\begin{equation}
\Pi(\mathbf{R},\mathbf{k};i\omega_{n}) = \{
[i\omega_{n}Z(\mathbf{R},\mathbf{q};i\omega_{n})^{2} -
[\xi_{\bk}^{2} + W^{2}(\mathbf{R},\mathbf{k};i\omega_{n})]^{2}
\}^{1/2}\;.
\end{equation}

Note that the strong correlation between electrons themselves can
give rise to an effective pairing. In this case, $v_{ee}$ is
negative. In principle, we cannot exclude the possibility  that
$v_{ee}$ also becomes inhomogeneous. Here we focus on the effect
from the electron coupled to local modes and will ignore the
contribution from $v_{ee}$. We mention equation for normal
self-energy correction for completeness.  As was pointed out,  we
will ignore $S$.

We adopt a six-parameter fit to the band structure used previously
for optimally doped Bi-2212 systems,~\cite{Norman:95} having the
form
\begin{eqnarray}
\xi_{\mathbf{k}}&=&-2t_{1} (\cos k_x + \cos k_y) -4t_{2} \cos k_{x}
\cos k_y \nonumber \\
&& -2t_{3} (\cos 2k_x + \cos 2k_y) \nonumber \\
&&-4t_{4} (\cos 2k_x \cos k_y + \cos k_x \cos 2k_y) \nonumber \\
&& -4 t_{5} \cos 2k_x \cos 2k_y - \mu \;,
\end{eqnarray}
where $t_1=1$, $t_{2}=-0.2749$, $t_{3}=0.0872$, $t_4=0.0938$,
$t_5=-0.0857$, and $\mu=-0.8772$. Unless specified explicitly, the
energy is measured in units of $t_1$ hereafter.

We use the method of Vidberg and Serene to first solve the above
coupled equations in the Matsubrara frequency space. On Matsubara
axis the quantities $S$ and $W$ are real.  We then do the analytical
continuation with the Pade approximation to covert them to the axis
of real frequency, on which they have a real and imaginary part.
Partly motivated by the ARPES experiments
\cite{Bogdanov00,Kaminski01,
Lanzara01,Johnson01,Zhou03,Kim03,Gromko03,Sato03,Cuk04,Zhou05}, we
take the averaged frequency of the local bosonic modes to be
$\langle \Omega_0 \rangle =0.3$ and the bare electron-bosonic mode
coupling constant $\langle g_0 \rangle =0.5$. The temperature is
chosen at $T=0.01$. To be illustrative, we first show the
calculation at a specific coarse-grained spatial point. Our
calculations show that $S$ is negligible and we will ingore it
hereafter.

\subsection{Relation between features of the gap and coupling constant and bosonic energy}

\begin{figure}[th]
\centerline{\psfig{file=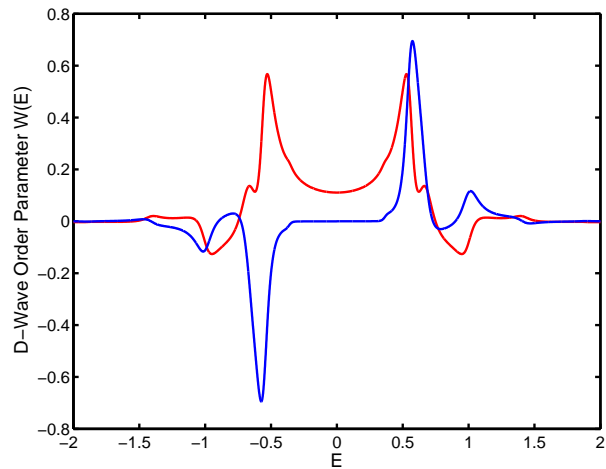,width=8cm}} \caption{The real (red
line) and imaginary (blue line) parts of the complex $d$-wave order
parameter as a function of energy. Here the bosonic mode frequency
$\Omega_{0}=0.3$ and the electron-mode coupling constant $g_0=0.5$.}
\label{FIG:W}
\end{figure}

\begin{figure}[th]
\centerline{\psfig{file=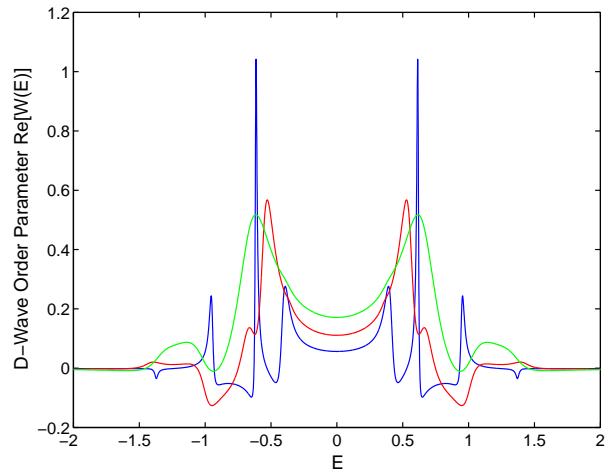,width=8cm}} \caption{The real part
of the complex $d$-wave order parameter as a function of energy for
various values of the electron-mode coupling constant $g_0=0.4$
(blue line), $0.5$ (red line), $0.6$ (green line). The bosonic mode
frequency is fixed at $\Omega_0=0.3$.} \label{FIG:W_cpl}
\end{figure}

\begin{figure}[th]
\centerline{\psfig{file=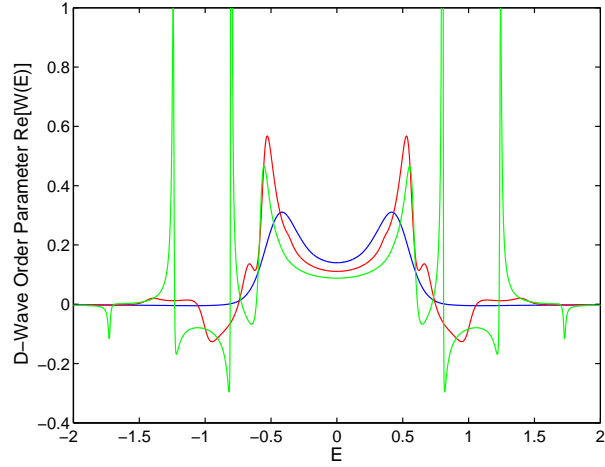,width=8cm}} \caption{The real part
of the complex $d$-wave order parameter as a function of energy for
various values of the mode frequency $\Omega_0=0.2$ (blue line),
$0.3$ (red line), $0.4$ (green line). The bare electron-mode
coupling constant is fixed at $g_0=0.5$.} \label{FIG:W_omg}
\end{figure}

In Fig.(\ref{FIG:W}) we can see that at low energies the real part
is constant and imaginary part rises only after energy exceeds the
boson energy $\Omega_0 = 0.3$. The first peak of $Re W$ curve and
shoulder on $Im W$ curve precisely correspond to the boson energy.
Features at higher energies correspond to the multiboson processes
and are at multiples of $\Omega_0$. In Fig.(\ref{FIG:W_cpl}) we
observe the evolution of the inelastic features as a function of
coupling constant. Gap function is changing substantially even for
fixed boson energy. Still there is always  a feature at boson energy
for all coupling constants. For larger energies and larger coupling
constants peaks in $Re W$ are getting broader. In
Fig.(\ref{FIG:W_omg}) the features in the the real part of gap
function for different values of boson energy are shown. Again the
first shoulder albeit of different intensity can be seen at energy
that exactly corresponds to the boson energy. Feature at $E = 0.2$
for blue curve is very broad, the one at $E = 0.3, 0.6$ can be seen
for boson energy $\Omega_0 = 0.3$ and finally for the green curve
one can see feature at $E = 0.4$. Features at substantially higher
energies  are likely to be numerical artifacts of our use of  Pade
approximations in analytic continuation.

\subsection{Correlation between inhomogeneous coupling constant and gap; anticorrelation between
bosonic mode energy and gap}

Next  we consider the effect of electronic inhomogeneity. For this
purpose, we consider two cases, a)  coupling constant has a spatial
distribution and b) local bosonic mode has a spatial distribution.
Both of these parameters can be position dependent at the same time
as we suspect is the case in real systems. Here  we want to
differentiate between  the effects coming from coupling constant and
effects coming from frequency variations. We assume that both
distributions are gaussians.

\begin{equation}
P(x) \propto \exp\biggl{[}\frac{(x-\langle x
\rangle)^{2}}{2\alpha^{2}}\biggl{]}\;,
\end{equation}
where $x$ represents $g_0$ or $\Omega_0$. Throughout the work, we
take $\alpha=0.3$.

 \begin{figure}[th]
\centerline{\psfig{file=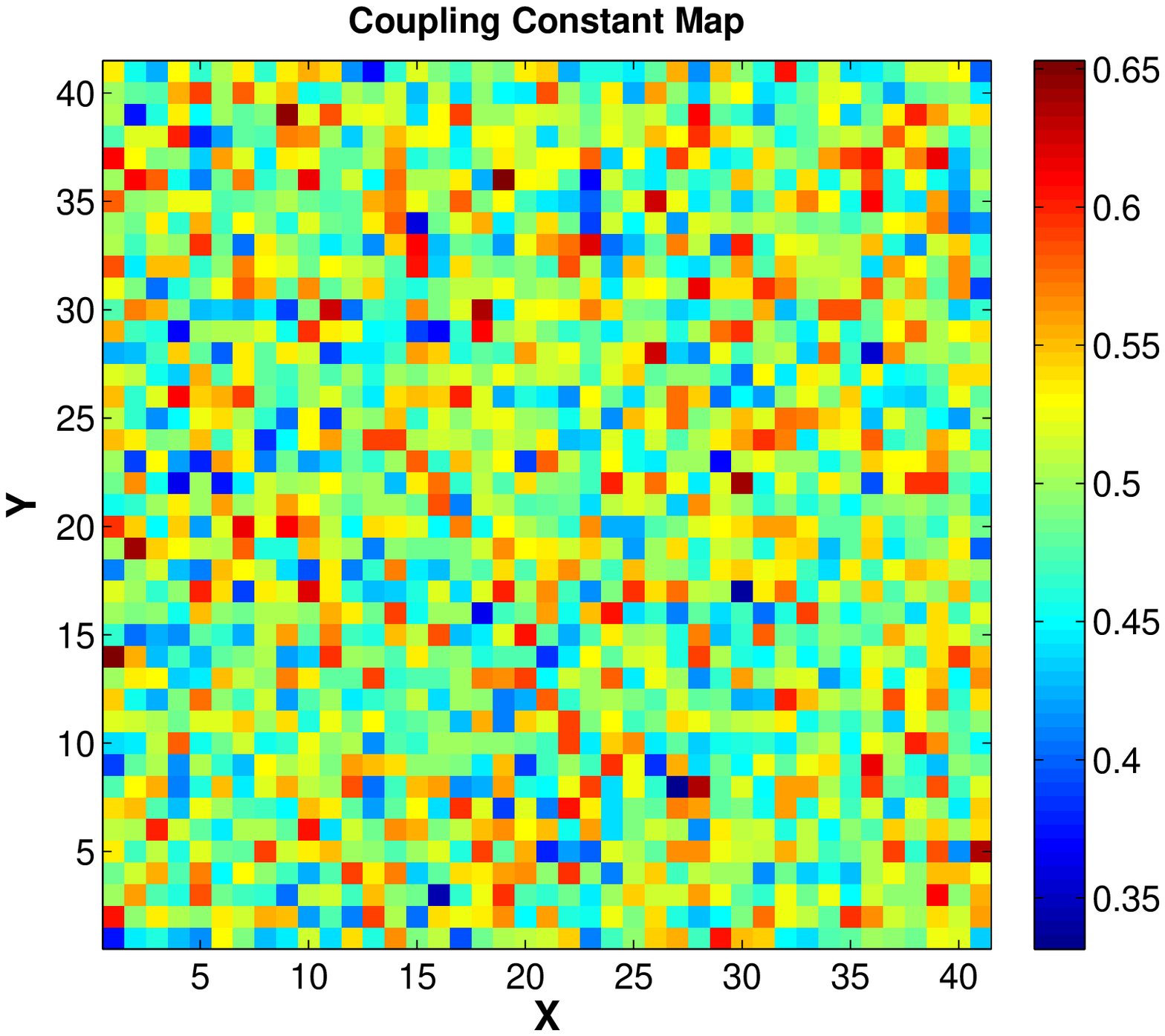,width=6cm}\psfig{file=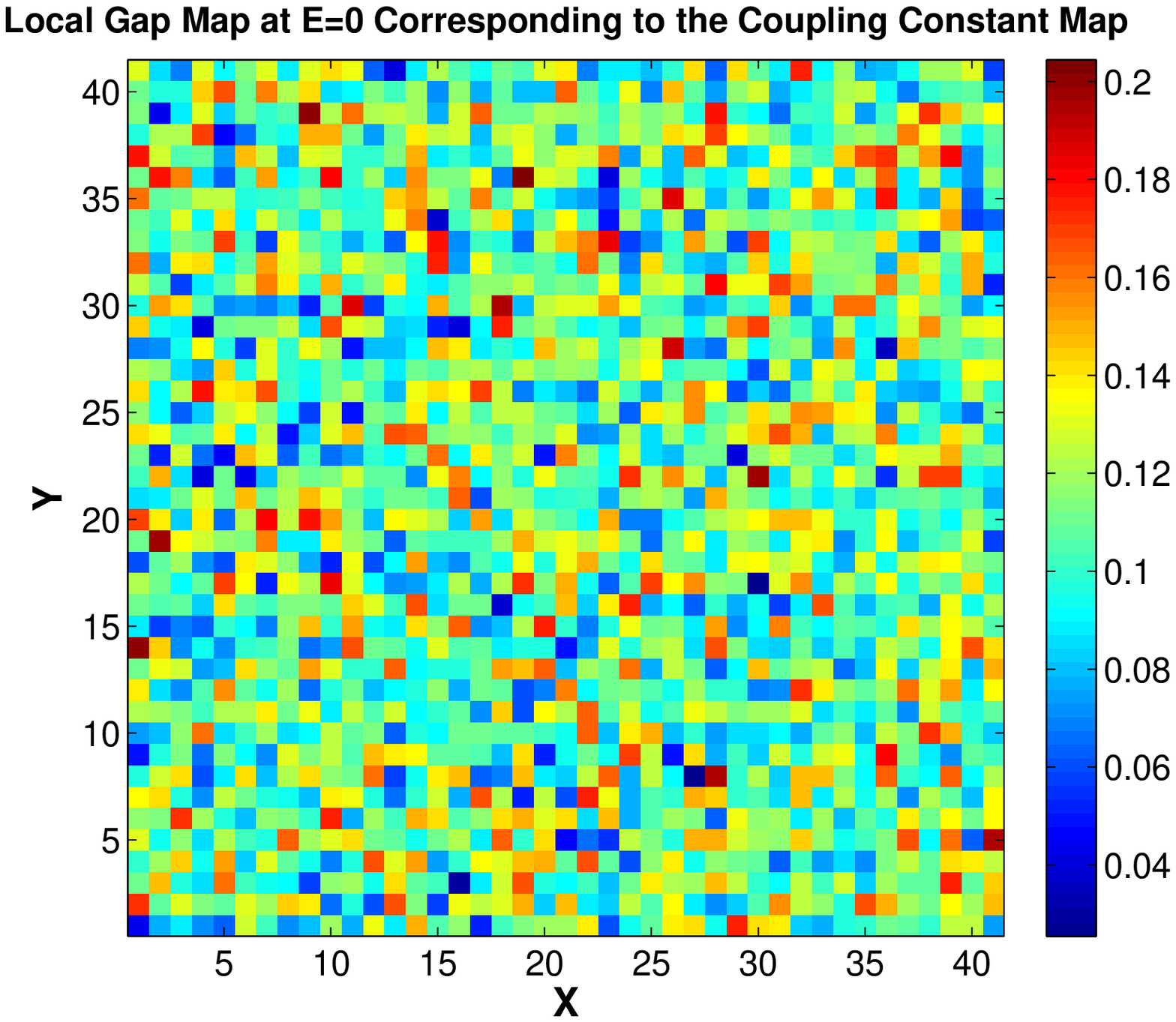,width=6cm}}
\caption{The spatial distributions of the bare coupling constant
(left panel) and the resultant $d$-wave order parameter at zero
energy. The bosonic mode frequency is fixed at $\Omega_0=0.3$. }
\label{FIG:cpl_gamma_map}
\end{figure}

\begin{figure}[th]
\centerline{\psfig{file=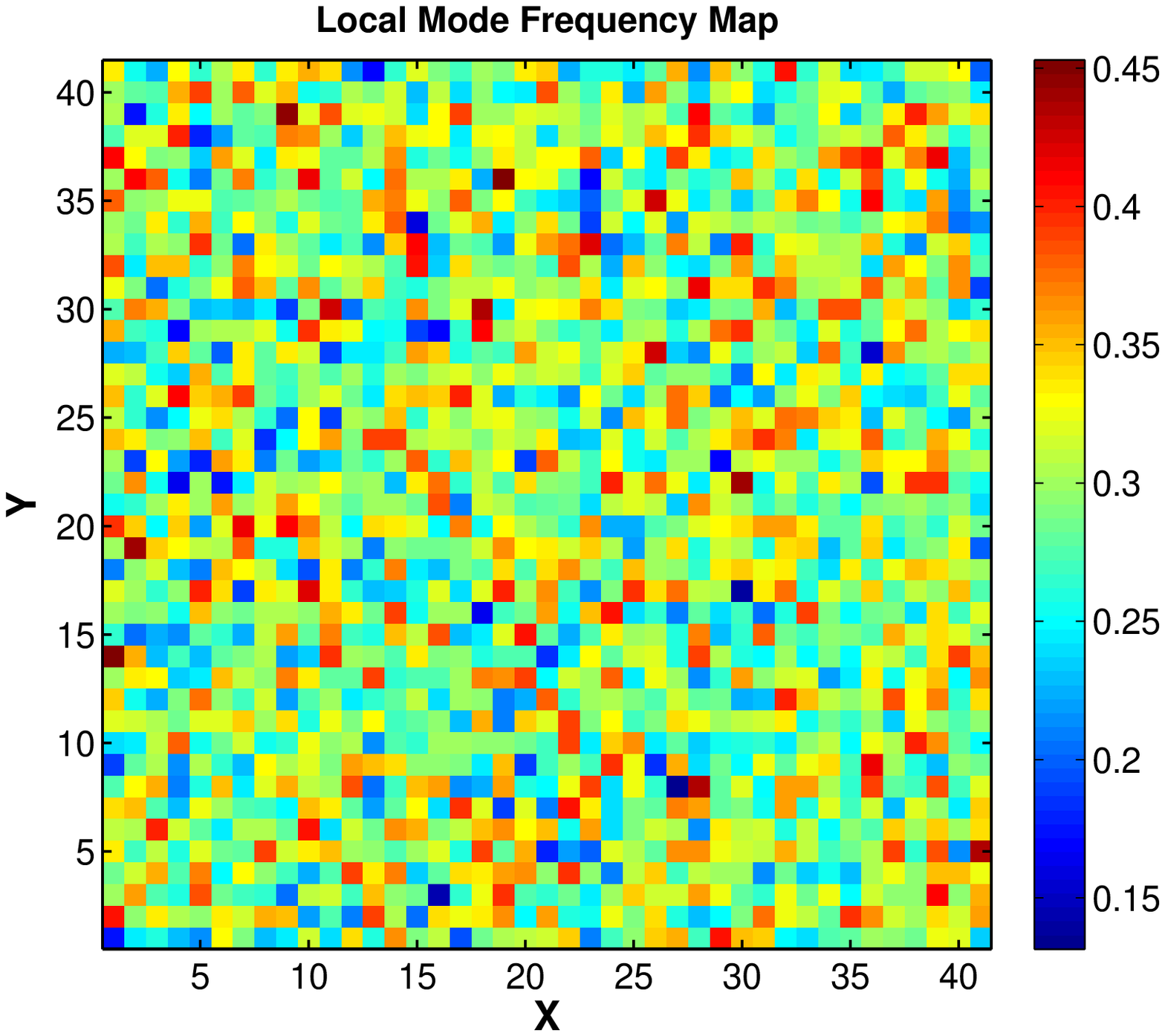,width=6cm}\psfig{file=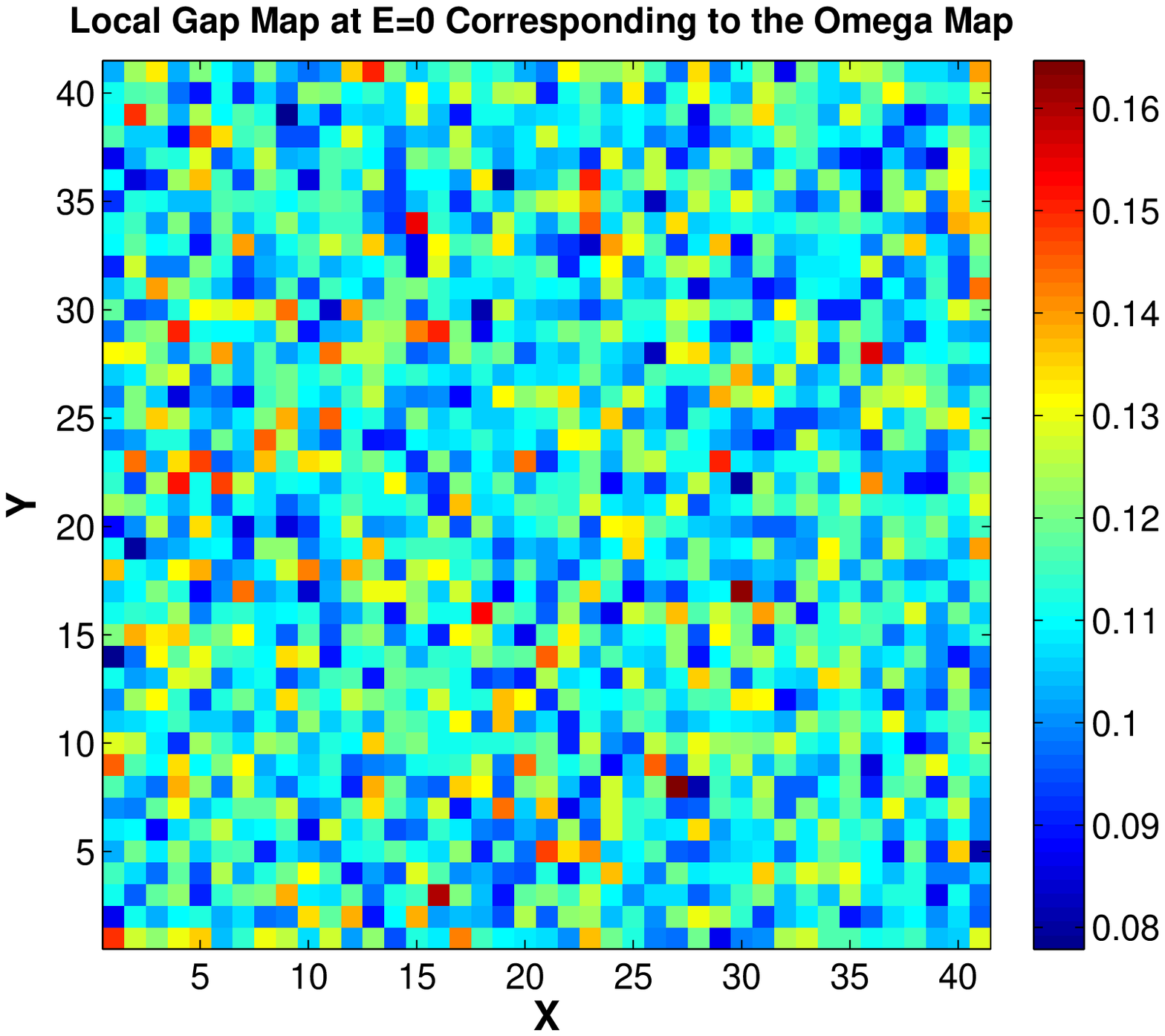,width=6cm}}
\caption{The spatial distributions of the mode frequency (left
panel) and the resultant $d$-wave order parameter at zero energy.
The bare electron-mode coupling constant is fixed at $g_0=0.5$.
Anticorrelation between the  mode frequency and gap magnitude is
clearly seen. Large  mode frequency regions (red spots in the left
panel) correlate with the small gap regions (dark spots in the right
panel). Anticorrelation for arbitrary coupling constant is thus
verified numerically.  In the weak coupling limit the
anticorrelation follows from Eq. (\ref{EQ:geff1},\ref{EQ:geff2}).}
\label{FIG:omega_gamma_map}
\end{figure}

In Fig(\ref{FIG:cpl_gamma_map}) we observe the direct correlation
between the strength of the coupling constant and $Re W(E = 0)$.
This direct correlation is natural and to be expected. In Fig.
(\ref{FIG:omega_gamma_map}) we find an {\em anticorrelation} between
the boson mode energy and  gap energy. The nature of this
anticorrelation follows from our assumption on the boson spectral
function that is peaked at one energy $\Omega_0$. Indeed from the
structure of the pairing kernel one can see that larger boson energy
at fixed $g$ would lead to {\em lower} effective coupling constant,
see Eq.(\ref{EQ:geff1}). Thus we conclude that the anticorrelation
is not a consequence of the weak coupling analysis but is present in
the full numerical solution of the self-consistent gap equations.
This anticorrelation is {\em directly} observed in the IETS STM
experiments.\cite{Jinho1}

\subsection{  LDOS map and the $d$-wave
order parameter map}

\begin{figure}
\centerline{\psfig{file=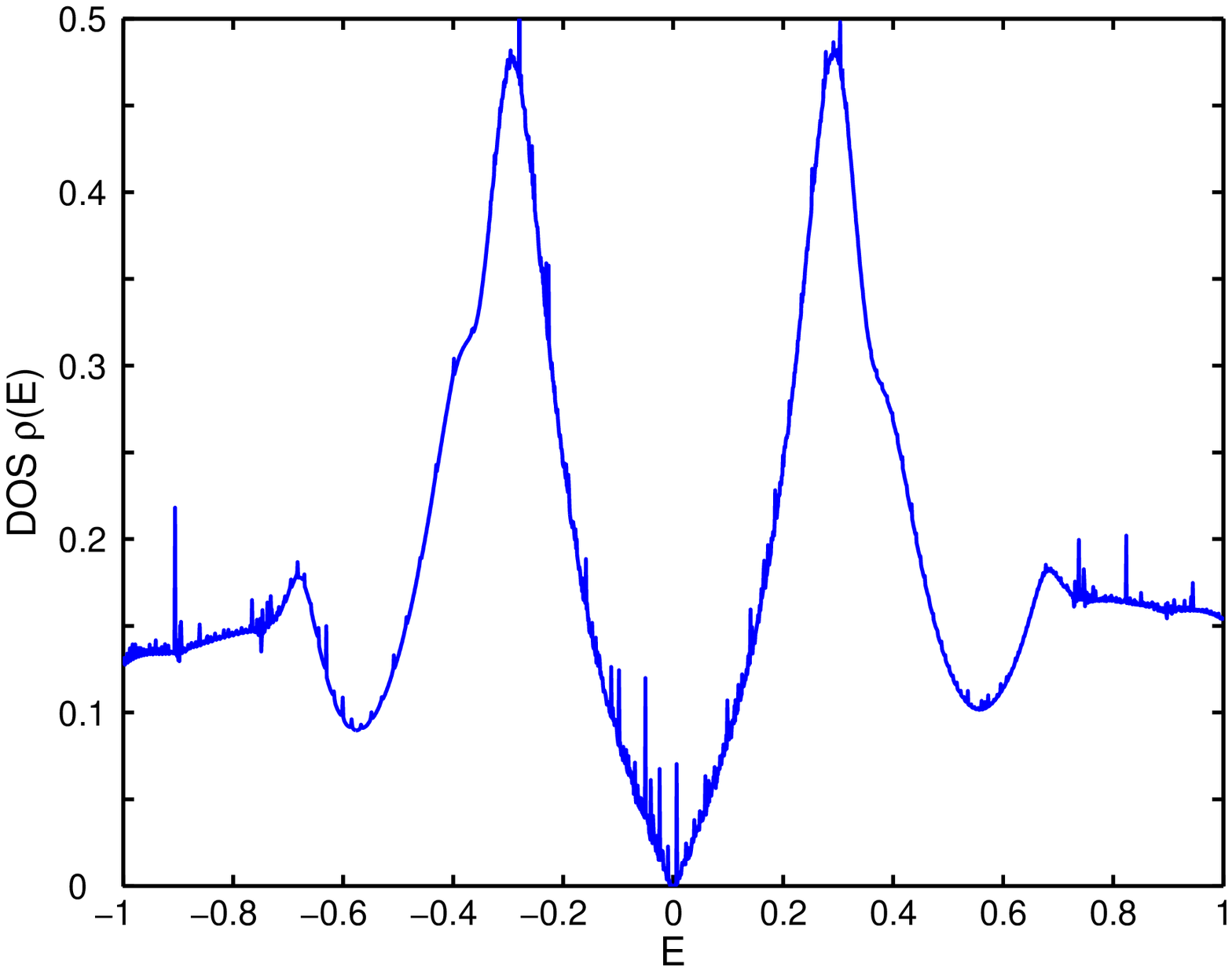,width=8cm}\psfig{file=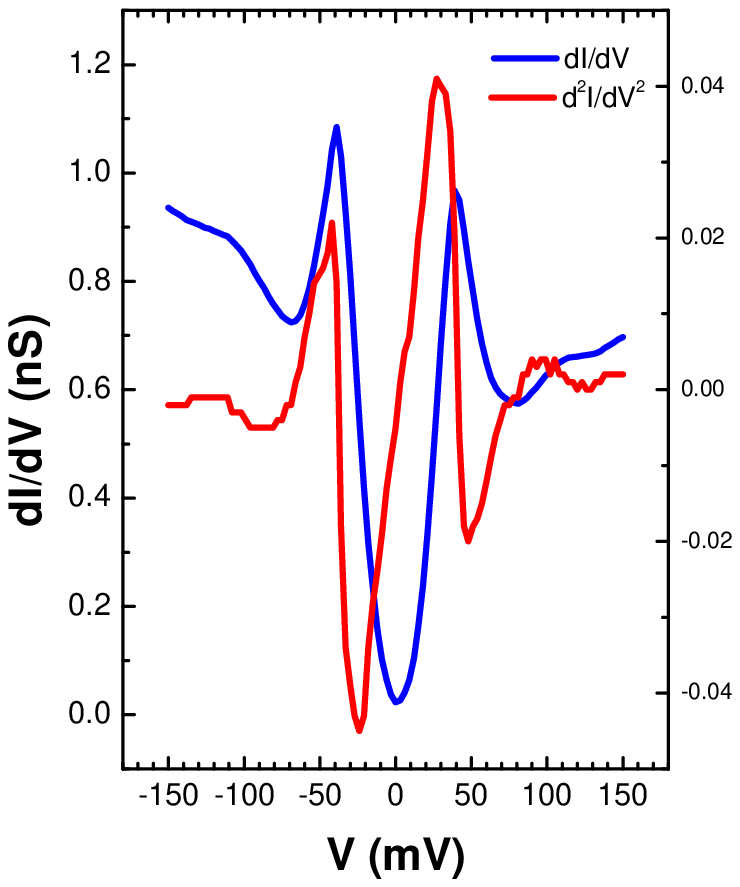,width=8cm}}
\caption{The density of states as a function of energy. Here the
bosonic mode frequency $\Omega_{B}=0.3$ and the electron-mode
coupling constant $g_0=0.5$. Also shown is an experimentally
measured $dI/dV$ and $d^2I/dV^2$  spectra at some typical point are
shown. The overall similarity of the spectrum with the one we have
calculated is suggestive that the features in the tunneling seen at
the large bias are indeed consistent with the strong coupling
features due to electron-boson interactions.} \label{FIG:dos}
\end{figure}

We have also calculated the local density of states (LDOS).

\begin{equation}
\rho(E) =\frac{2}{N_{L}\pi} \sum_{\mathbf{k}} \vert \text{Im}
\mathcal{G}(\mathbf{R},\mathbf{k};E+i0^{+})\vert \;,
\end{equation}
which correspond to the local differential tunneling conductance as
measured by the STM experiments. Fig.(\ref{FIG:dos}) shows the local
density of states as a function of energy at a selected spatial
point corresponding to Fig.1. For comparison we also show typical
experimental data for STM tunneling density of states and its
derivative, Fig(\ref{FIG:dos})

\begin{figure}[th]
\centerline{\psfig{file=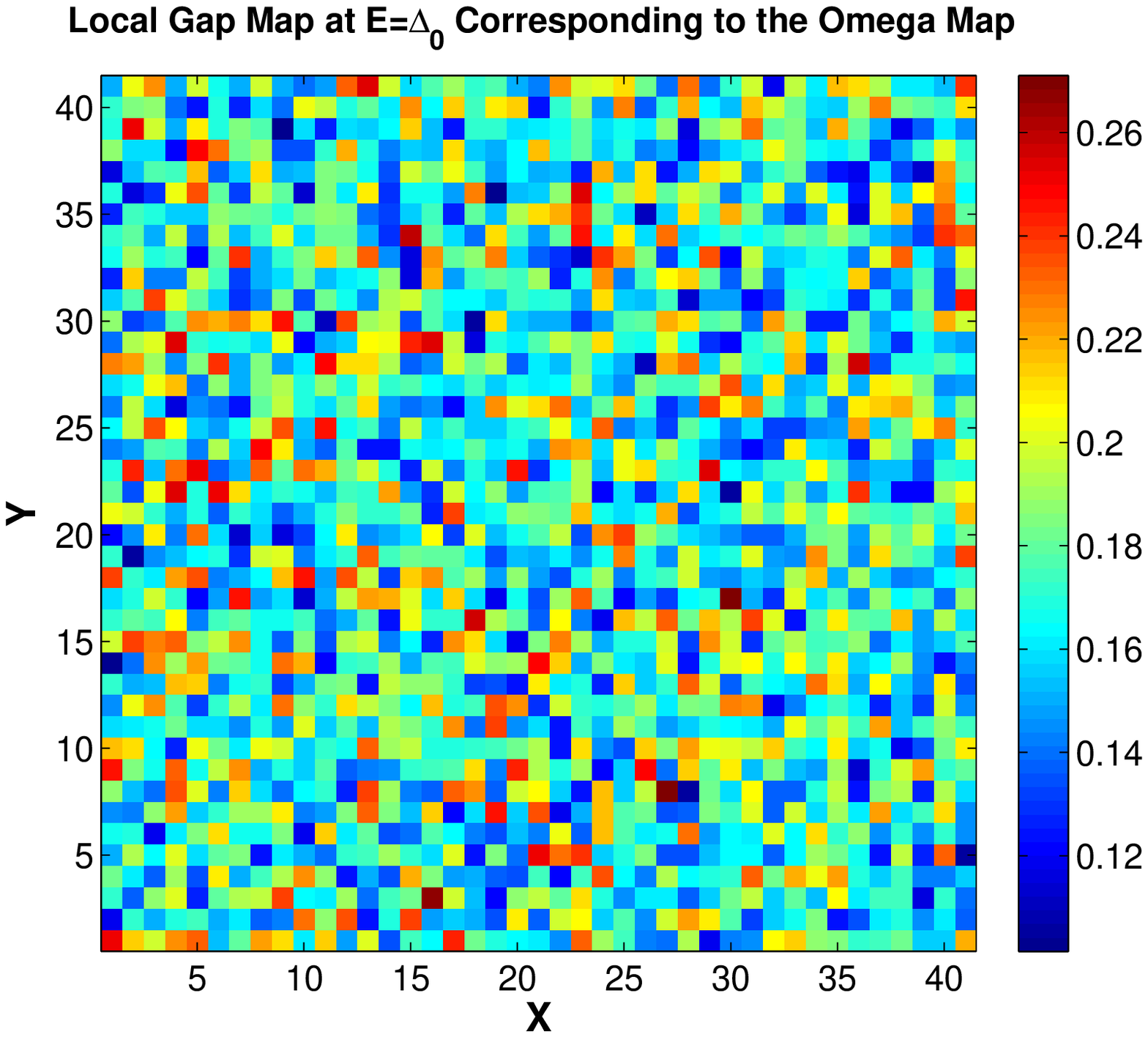,width=6cm}\psfig{file=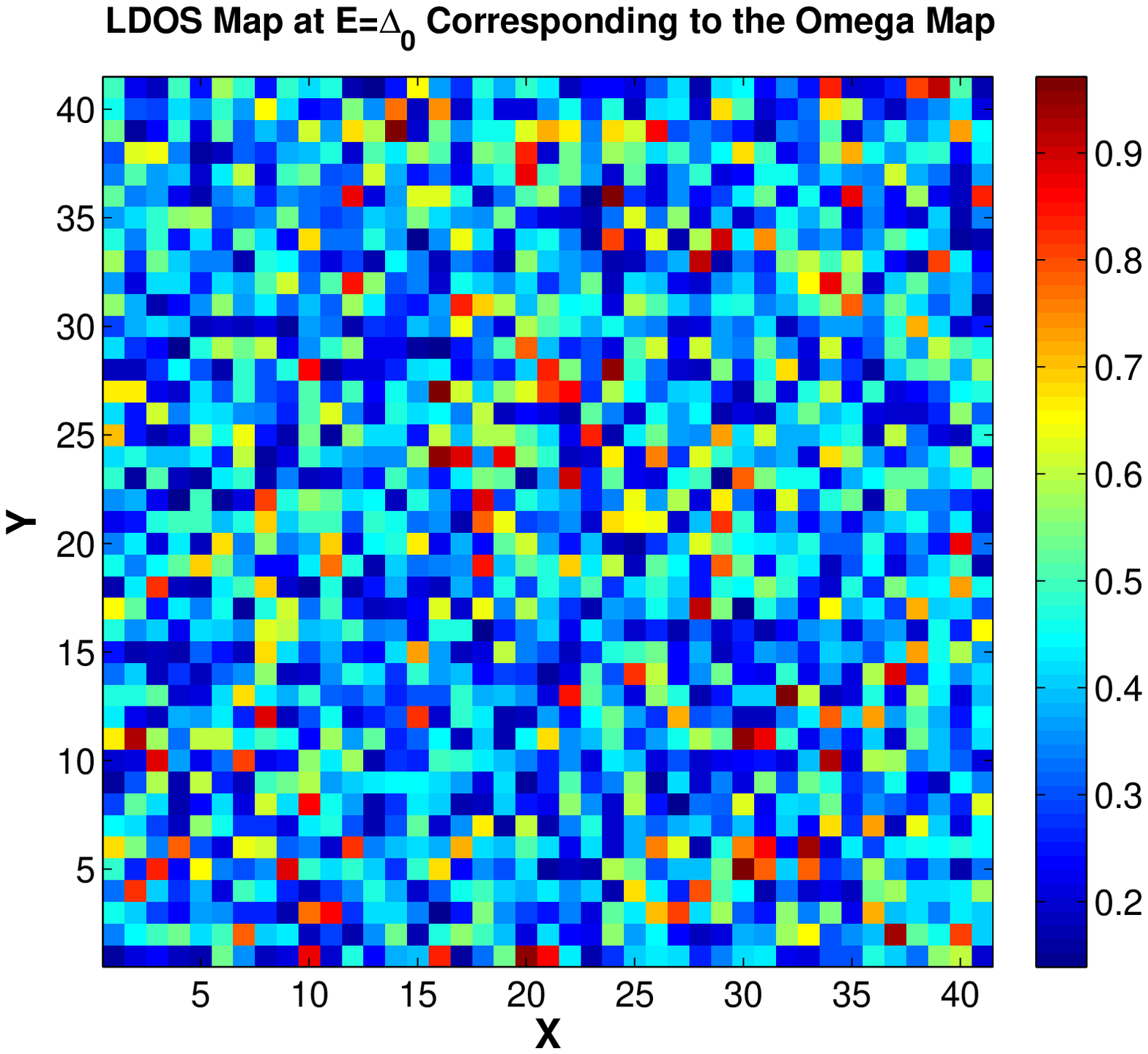,width=6cm}}
\caption{The spatial distributions of the real part of the $d$-wave
order parameter at $E=\Delta_g$ (left panel) and the local density
of states at the same energy. The bare electron-mode coupling
constant is fixed at $g_0=0.5$.} \label{FIG:gamma_dos_map}
\end{figure}

We also calculated  spatial image of the LDOS at the energy
$\Delta_{g}$. In the strong-coupling theory, the $d$-wave order
parameter is energy dependent. To demonstrate the distribution of
gap inhomogeneity from the LDOS distribution at the gap edge
$\Delta_g$, we should use the $d$-wave order parameter at fixed
energy, say  $E = <\Delta>$. Fig(\ref{FIG:gamma_dos_map}) shows
spatial distribution of the LDOS for the case of an inhomogeneous
distribution of local bosonic modes.

\section{Isotope Effect in Inhomogeneous Superconductor}

One of the most powerful tools to investigate the role of phonons in
the pairing is to study isotope effect. In the case of conventional
supercoductors it was found that isotope subsitution of the lattice
atoms would change the transiton temperature thus directly
indicating that lattice is involved in pairing. By changing the mass
of the ions from M to M' critical temperature $T_c \sim M^{-\alpha}$
would change by $T_c'/T_c = (M/M')^{\alpha}$.

First consider conventional homogeneous superconductors. In
conventional superconductors often $\alpha \simeq 1/2$ is observed.
This result follows the observation that the effective coupling
constant in standard BCS formalism is independent of mass
M.\cite{Scalapino:69}  Simple arguments show that effective coupling
constant in homogeneous case $g_{eff} = \frac{C}{M \omega^2_{av}}$,
where $C$ is a constant, $M$ is the ion mass and $\omega^2_{av}$ is
the average phonon frequency squared. Since $\omega_{av} \sim
M^{-1/2}$ for phonon spectrum regardless of the detailed shape,  one
finds that the coupling constant in this case is independent of mass
$M$. Therefore the only effect of the isotope substitution is on the
change of the phonon spectrum and the cutoff frequency that is in
the prefactor in BCS equation for \beqa T_c = \Omega_0 \exp
(-\frac{1}{N_0g_{eff}}) \label{EQ:isotope1} \eeqa

 We thus find the conventional  exponent is   $\alpha = 1/2$ that is set
solely by the prefactor $\Omega_0$ within the  BCS theory. Situation
we consider is very different. As we pointed out, $g_{eff}(\bR)$ is
made from two random {\em independent} quantities, $g(\bR)$ and
$\Omega_0(\bR)$. This will lead to a very different isotope effect
in this random superconductor.

Standard isotope substitution experiment in cuprates is replacement
of $^{16}O$ by $^{18}O$. The changes in $T_c$ produced by this
substitution are small, $\alpha$ is small and depend on doping
levels of the samples. In the optimally doped samples $\alpha$ is
essentially zero.

Situation for inhomogeneous superconductors is qualitatively
different from conventional BCS case. Inhomogeneous superconductor
is characterized by {\em two rather than one} parameter that enters
into the gap equation Eq.(\ref{EQ:local4}): one is coupling constant
$g(\bR)$ and another one is a local boson frequency $\Omega_0(\bR)$.
In principle, both random variable will change upon isotope
substitution. On general grounds,  isotopic substitution would
change the local environement as it affects both in-plane and out
of-plane oxygen atoms. Hence, we argue that {\em both the coupling
constant and phonon frequency are affected} by isotope substitution.
It would mean therefore that {\em both prefactor and coupling
constant in the exponent are changed} in Eq.(\ref{EQ:local9}) upon
$^{16}O$ to $^{18}O$ substitution.  In addition, for the
inhomogeneous superconductor one has to differentiate between the
isotope shift of critical temperature of a  sample $T_c$ and the
isotope shift of the gap $\Delta(\bR)$. Here we do not address the
net shift of $T_c$ as it is determined by phase fluctuations at
higher temperatures. Instead we focus on shift of local gap
$\Delta(\bR)$ or average gap $<\Delta(\bR)>$.

\beqa \Delta(\bR) = \Omega_0(\bR) \exp(-\frac{1}{N_0 g_{eff}(\bR)})
\label{EQ:isotope2} \eeqa

 Again for simplicity we will
take a weak coupling limit of inhomogeneous Eliashberg equations.
Let us take an ''average" of the equation as an approximate way to
discuss the average shifts in $\Omega_0(\bR)$ and in
$N_0g_{eff}(\bR)$,  Eq.(\ref{EQ:isotope2})

\beqa \langle\Delta(\bR)\rangle = \langle\Omega_0(\bR)\rangle
\exp(-\frac{1}{\langle N_0 g_{eff}(\bR)\rangle}) \label{EQ:isotope3}
\eeqa

It is known \cite{Jinho1} that the average frequency $\Omega_0 =
\langle\Omega_0(\bR)\rangle$ shifts from 52 meV to 48 meV upon
$^{16}O$ by $^{18}O$ substitution.  On the other hand this shift in
$\Omega_0$ can be offset by a shift in average $g_{eff}$ thus
producing an  zero isotope effect that is very different from BCS
exponent $\alpha_{BCS} = 1/2$. To illustrate how one gets near zero
isotope shift,  we take that effective coupling constant shifts by

\beqa \ln (\frac {\Omega_0}{\Omega'_0}) = \frac{1}{N_0g_{eff}} -
\frac{1}{N_0g'_{eff}} \label{isotope4} \eeqa

This shift of the effective coupling constant could offset the shift
of the prefactor in Eq.(\ref{EQ:isotope3}). The net isotope effect
will be determined by the combined isotope shift of the boson mode
energy and effective coupling constant. They can mutually cancel
each other making net effect small, as we suspect is the case near
optimal doping. Alternatively, both effects   can   add up to
produce large isotope shift. Thus by knowing only the shift of boson
energy is not sufficient to address the net isotope effect of the
gap.

\begin{figure}[th]
\centerline{\psfig{file=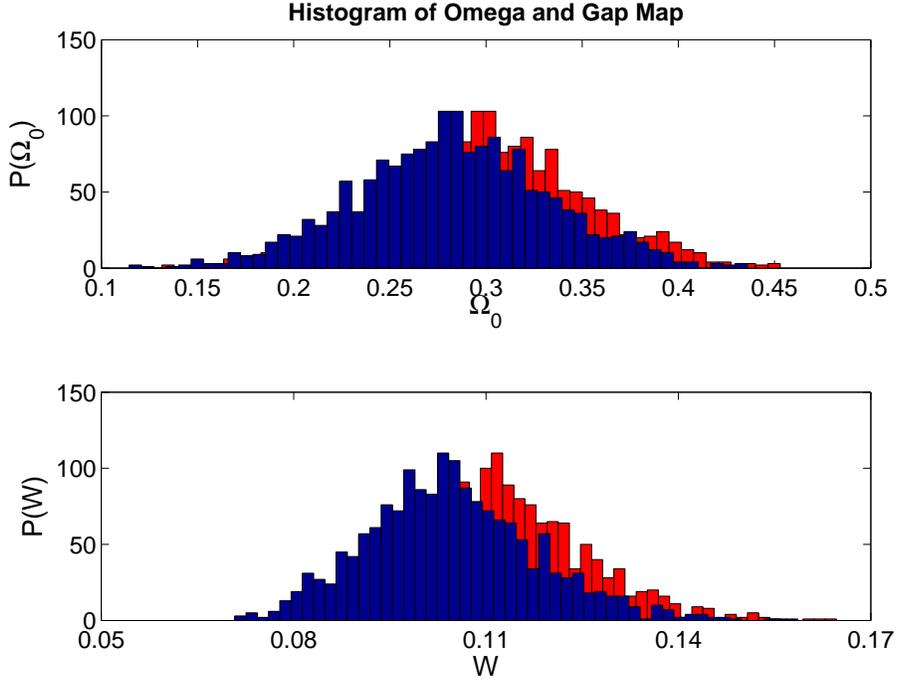,width=12cm}} \caption{ Upper
panel: histograms of (the boson energy $\Omega_0$ (upper panel)  for
$O^{16}$ (red) and $O^{18}$ (blue) are shown assuming $6\% $ isotope
shift. Lower panel: histograms of the calculated gap distribution
$W$ for these two boson mode distribution calculated with
simultaneous coupling constant shift. We assumed that coupling
relative constant shift is negative $4\% $. This turns to be
sufficient to produce net negative gap shift.} \label{FIG:Isoshift}
\end{figure}

To illustrate this point we have preformed "numerical isotope shift"
experiment within our model, see Fig(\ref{FIG:Isoshift}).  In order
to model the effect of oxygen subtitution we  change random
distribution $P(\Omega_0)$ of  the boson energy $\Omega_0$. We
assumed that changing $^{16}O$ to $^{18}O$ would shift gaussian
distribution with the mean values of boson energy $\langle \Omega_0
\rangle_{O16} = \overline{\Omega} = 0.3$ to $\langle \Omega_0
\rangle_{O18} = \overline{\Omega} (1-6\%)$ ( upper panel). At the
same time coupling constant $g$ can also change upon isotope
substitution.  We consider the shift in the coupling constant $g$
which here is taken to be constant at the same time (lower panel).
For $^{16}O$ we take $g = g_0 = 0.5$ and for $^{18}O$ we use
$g_0(1-4\%)$. We find that negative shift of boson energy by $6\%$
can be offset by the shift of the coupling constat $g_0$ and the net
effect would be negative shift of the gap $\langle W \rangle$. To
address the net isotope effect it is necessary to measure
independently both boson energy and coupling constat for $^{16}O$
and $^{18}O$. At moment there is  no independent experimental
measurement of the coupling constant we are aware of.

\begin{figure}[th]
\centerline{\psfig{file=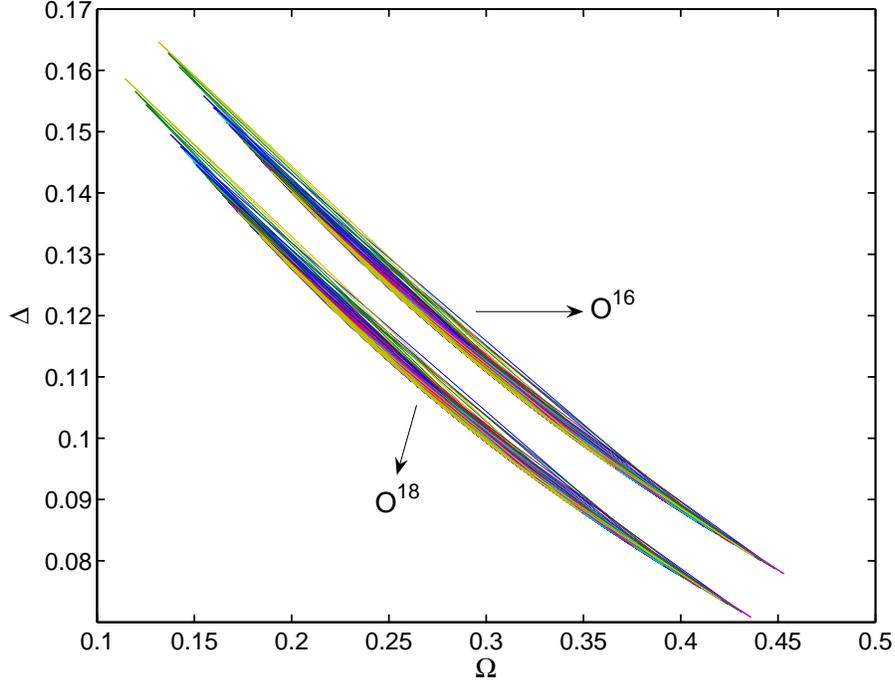,width=12cm}} \caption{From
the real space maps for mode energy $\Omega_0$ and gap as a function
of position we can extract the statistical cross-correlation of
these two quantities. We find overall negative concave-shaped
$\Delta$-$\Omega_0$ cross correlation for the cases of $O^{16}$ and
$O^{18}$. Isotope shift of mode energy to lower energies is clearly
seen on this plot. The average gap value is affected by the shift in
the coupling constant and in mode energy. }\label{FIG:Isoshift2}
\end{figure}

For completeness we also  present the correlation between the gap
and the mode energy $\Omega_0$, Fig(\ref{FIG:Isoshift2}). The
negative correlation between the gap and mode energy is  clearly
seen for both $O^{16}$ and $O^{18}$ isotopes.

\section{Conclusion}

In this paper we use strong coupling model for boson mediated d-wave
pairing for inhomogeneous superconductor.  To model  inhomogeneous
superconductor we consider a coarse grain model with the typical
size of the patch on the order of superconducting coherence length
$\xi \sim 20-50 \AA$. We use patch dependent pairing interaction due
to disordered pairing boson with patch-dependent coupling constant
$g(\bR)$ and boson energy $\Omega_0(\bR)$. This local pairing
produces local superconducting gap as a selfconsistent solution  of
Eliashberg equations patch by patch.

We argue that any inhomogeneous theory of strong coupling theory of
pairing has to involve at least {\em two} independent quantities
that characterize electron-boson interaction: coupling constant and
boson energy. Reduction of the pairing theory to an inhomogeneous
BCS pairing model does not allow one to distinguish a relative role
of the coupling constant $g(\bR)$ vs $\Omega_0(\bR)$. In BCS-like
analysis one deals with the single effective coupling $g_{eff}(\bR)
$ that is combination of $g(\bR)$ and $\Omega_0(\bR)$,
Eq.(\ref{EQ:geff2}).

We use the local coupling constant and boson energy drawn randomly
with the gaussian distributions and calculate local gap function
$W(\omega, \bR) = Re W(\omega, \bR) + i Im W(\omega, \bR)$. This gap
map is then used to compare correlations between $W(\bR)$, $g(\bR)$
and $\Omega_0(\bR)$. We also calculate the local tunneling density
of states that is consistent with the observed by STM density of
states.

Numerically we find a direct positive correlation between the gap
map $W(\bR)$ and boson coupling constant map. This is not
surprising: the larger the coupling the larger is gap function. We
also find an interesting and surprising result that there is a {\em
negative} correlation between gap function scale and boson mode
energy $\Omega_0$. We give a simple interpretation of this negative
correlation in the case of the weak coupling analysis. We find that
effective BCS coupling constant {\em is inversely proportional } to
the boson energy, Eq.(\ref{EQ:geff2}).  Because effective coupling
constant is in the exponent of a gap solution Eq.(\ref{EQ:local9}),
its dependence on $\Omega_0(\bR)$ is more important than the
dependence on frequency in the prefactor. Larger frequency boson is
less effective in inducing pairing for fixed $g(\bR)$ in our model.
This result is consistent with the experimental observation of the
IETS signal, where {\em anticorrelation} between gap and boson mode
energy was observed.  \cite{Jinho1}

Exact nature of the boson is not important for our analysis except
when we discuss isotope effect. Then  we expicitly assume that boson
is a lattice mode and its energy has on average an isotope shift
consistent with the isotope shift for phonons due to $^{16}O$ to
$^{18}O$ substitution.

 We  also consider
isotope effect. We find that in order to correctly address the full
isotope effect one again would need to have assess changes in the
coupling constant $g(\bR)$ and $\Omega_0(\bR)$ as a result of
$^{16}O$ to $^{18}O$ isotope substitution. Both shift in the boson
frequency and coupling constant contribute to the net isotope effect
and we find that small, on the order of $\sim 4 \%$,  changes of the
coupling constant can completely offset the isotope shift of the gap
function $W(\bR)$ caused by the standard isotope effect for the
lattice mode $\Omega_0(\bR)$.

We therefore find that in order to understand the isotope effect in
d-wave superconductors we would need independent measurements of the
coupling constant map and boson mode energy map for different
isotopes. The whole notion that a single isotope exponent can
characterise the spatially modulated superconducting state as is the
case of high-Tc materials seem to be too simplistic to address the
real situation. It appears one can not make an evaluation on the
importance of the lattice effects in high-Tc superconductors based
on a shift of a critical temperature without addressing the changes
of the gap,  boson modes and coupling constant.

To address the effects of spatial inhomogeneity of tunneling IETS
spectra we focus on electron-boson coupling, ignoring
electron-electron interaction part that will not produce IETS
features. Electron-boson interaction  is only one contribution to
the pairing interactions. Pairing in high-$T_c$ materials is likely
a result of interplay between strong electron-electron correlations
and electron-boson interaction. The realistic magnitude of pairing
interaction and how large the transition temperature would be by
assuming only electron-boson pairing is an interesting question. We
leave this question for a separate investigation.

 \acknowledgments

We thank Ar. Abanov, A. Chubukov, J.C. Davis,  D.-H. Lee, Jinho Lee,
N. Nagaosa, M. R. Norman, D. J. Scalapino, and Z. X. Shen for very
useful discussions. This work was supported by the US DOE.


\begin{thebibliography}{99}

\bibitem{Eliashberg:60} G. M. Eliashberg, Sov. Phys. JETP {\bf 11}, 696 (1960).

\bibitem{McMillan:65} W. L. McMillan and J. M.  Rowell, Phys. Rev. Lett. {\bf 14}, 108
(1965);  W. L. McMillan and J. M. Rowell, in {\em
Superconductivity},
 Vol. 1, edited by R. D. Parks (Dekker, New York, 1969).

\bibitem{Scalapino:69} D. J. Scalapino, in {\em Superconductivity}, Vol. 1, edited by R. D. Parks
(Dekker, New York, 1969).

\bibitem{Carbotte:90} J. P. Carbotte,
Rev. Mod. Phys. {\bf 62}, 1027 (1990).

\bibitem{Giaever:60} I. Giaever, Phys. Rev. Lett. {\bf 5}, 464
(1960); J. C. Fisher and I. Giaever, J. Appl. Phys. {\bf 32}, 172
(1961); I. Giaever, Phys. Rev. Lett. {\bf 5}, 146 (1966).

\bibitem{Huang89} Q. Huang, J. F. Zasadzinski, K. E. Gray, J. Z.
Liu, and H. Claus, Phys. Rev. B {\bf 40}, 9366 (1989).

\bibitem{Renner95} Ch. Renner and \O. Fischer, Phys. Rev. B {\bf
51}, 9208 (1995).

\bibitem{Renner96} Ch. Renner, B. Revaz, J.-Y. Genoud, and \O.
Fischer, J. Low Temp. Phys. {\bf 105}, 1083 (1996).

\bibitem{DeWilde98} Y. DeWilde, N. Miyakawa, P. Guptasarma, M.
Iavarone, L. Ozyuzer, J. F. Zasadzinski, P. Romano, D. G. Hinks, C.
Kendziora, G. W. Crabtree, and K. E. Gray, Phys. Rev. Lett. {\bf
80}, 153 (1998).

\bibitem{Mandrus91} D. Mandrus, L. Forro, D. Koller, and L. Mihaly,
Nature {\bf 351}, 460 (1991).

\bibitem{Yurgens99} A. Yurgens, D. Winkler, T. Claeson, S.-J. Hwang,
and J.-H. Choy, Int. J. Mod. Phys. B {\bf 29-31}, 3758 (1999).

\bibitem{Zasadzinski00} J.F. Zasadzinski, L. Ozyuzer, N. Miyakawa,
D. G. Hinks, K. E. Gray, Physica C {\bf 341-348}, 867 (2000).

\bibitem{Zasadzinski01} J. F. Zasadzinski, L. Ozyuzer, N. Miyakawa,
K. E. Gray, D. G. Hinks, and C. Kendziora, Phys. Rev. Lett. {\bf
87}, 067005 (2001).

\bibitem{Dessau91} D. S. Dessau, B. O. Wells, Z.­X. Shen, W. E.
Spicer, A. J. Arko, R. S. List, D. B. Mitzi, A. Kapitulnik, Phys.
Rev. Lett. {\bf 66}, 2160 (1991).

\bibitem{Ding96} H. Ding, A. F. Bellman, J. C. Campuzano,
M. Randeria, M. R. Norman, T. Yokoya, T. Takahashi, H.
Katayama-Yoshida, T. Mochiku, K. Kadowaki, G. Jennings, and G. P.
Brivio, Phys. Rev. Lett. {\bf 76}, 1533 (1996).

\bibitem{Shen93}  Z.X. Shen et al., Phys. Rev. Lett., {\bf 70}, 1553 (1993).


\bibitem{Campuzano99} J. C. Campuzano, H. Ding, M. R. Norman,
H. M. Fretwell, M. Randeria, A. Kaminski, J. Mesot, T. Takeuchi, T.
Sato, T. Yokoya, T. Takahashi, T. Mochiku, K. Kadowaki, P.
Guptasarma, D. G. Hinks, Z. Konstantinovic, Z. Z. Li, and H. Raffy
Phys. Rev. Lett. {\bf 83}, 3709 (1999).

\bibitem{Bogdanov00} P. V. Bogdanov, A. Lanzara, S. A. Kellar, X. J. Zhou,
E. D. Lu, W. J. Zheng, G. Gu, J.-I. Shimoyama, K. Kishio, H. Ikeda,
R. Yoshizaki, Z. Hussain, and Z. X. Shen, Phys. Rev. Lett. {\bf 85},
2581 (2000).

\bibitem{Kaminski01} A. Kaminski, M. Randeria, J. C. Campuzano, M. R.
Norman, H. Fretwell, J. Mesot, T. Sato, T. Takahashi, and K.
Kadowaki, Phys. Rev. Lett. {\bf 86}, 1070 (2001).

\bibitem{Lanzara01} A. Lanzara, P. V. Bogdanov, X. J. Zhou, S. A. Kellar,
D. L. Feng, E. D. Lu, T. Yoshida, H. Eisaki, A. Fujimori, K. Kishio,
J.-I. Shimoyama, T. Noda, S. Uchida, Z. Hussain, Z.-X. Shen, Nature
{\bf 412}, 510 (2001); Gweon, G.-H. et al.   Nature {\bf 430},
187–190 (2004).

\bibitem{Johnson01} P. D. Johnson, T. Valla, A. V. Fedorov,
Z. Yusof, B. O. Wells, Q. Li, A. R. Moodenbaugh, G. D. Gu, N.
Koshizuka, C. Kendziora, Sha Jian, and D. G. Hinks, Phys. Rev. Lett.
{\bf 87}, 177007 (2001).


\bibitem{Zhou03} X. J. Zhou, T. Yoshida, A. Lanzara, P. V. Bogdanov,
S. A. Kellar, K. M. Shen, W. L. Yang, F. Ronning, T. Sasagawa, T.
Kakeshita, T. Noda, H. Eisaki, S. Uchida, C. T. Lin, F. Zhou, J. W.
Xiong, W. X. Ti, Z. X. Zhao, A. Fujimori, Z. Hussain, Z.-X. Shen,
Nature {\bf 423}, 398 (2003).

\bibitem{Kim03} T. K. Kim, A. A. Kordyuk,
S. V. Borisenko, A. Koitzsch, M. Knupfer, H. Berger, and J. Fink,
Phys. Rev. Lett. {\bf 91}, 167002 (2003).

\bibitem{Gromko03} A. D. Gromko, A. V. Fedorov, Y.-D. Chuang,
J. D. Koralek, Y. Aiura, Y. Yamaguchi, K. Oka, Yoichi Ando, and D.
S. Dessau, Phys. Rev. B {\bf 68}, 174520 (2003).

\bibitem{Sato03} T. Sato, H. Matsui, T. Takahashi, H. Ding, H.-B. Yang,
S.-C. Wang, T. Fujii, T. Watanabe, A. Matsuda, T. Terashima, and K.
Kadowaki, Phys. Rev. Lett. {\bf 91}, 157003 (2003).

\bibitem{Cuk04}  T. Cuk, F. Baumberger, D. H. Lu, N. Ingle,
X. J. Zhou, H. Eisaki, N. Kaneko, Z. Hussain, T. P. Devereaux, N.
Nagaosa, and Z.-X. Shen, Phys. Rev. Lett. {\bf 93}, 117003 (2004).


\bibitem{Zhou05} X.J. Zhou et al., Phys. Rev. Lett. {\bf 95}, 117001
(2005).

\bibitem{Dahm96} T. Dahm, D. Manske, D. Fay, and L. Tewordt, Phys.
Rev. B {\bf 54}, 12006 (1996); T. Dahm, D. Manske, and L. Tewordt,
Phys. Rev. B {\bf 58}, 12454 (1998).

\bibitem{Shen97} Z. X. Shen and J. R. Schrieffer, Phys. Rev. Lett. {\bf
78}, 1771 (1997).

\bibitem{Norman97} M. R. Norman, H. Ding, J. C. Campuzano,
T. Takeuchi, M. Randeria, T. Yokoya, T. Takahashi, T. Mochiku, and
K. Kadowaki, Phys. Rev. Lett. {\bf 79}, 3506 (1997).

\bibitem{Norman98}M. R. Norman and H. Ding, Phys. Rev. B {\bf 57}, R11089
(1998).

\bibitem{Abanov99} Ar. Abanov and A. V. Chubukov, Phys. Rev. Lett.
{\bf 83}, 1652 (1999); Phys. Rev. B {\bf 61}, R9241 (2000).

\bibitem{Eschrig00} M. Eschrig and M. R. Norman, Phys. Rev. Lett.
{\bf 85}, 3261 (2000); Phys. Rev. B {\bf 67}, 144503 (2003).

\bibitem{Norman01} M. R. Norman, M. Eschrig, A. Kaminski, and J. C.
Campuzano, Phys. Rev. B {\bf 64}, 184508 (2001).

\bibitem{Manske01} D. Manske, I. Eremin, and K. H. Bennemann, Phys.
Rev. lett. {\bf 87}, 177005 (2001).

\bibitem{Abanov03} Ar. Abanov, A. V. Chubukov, and J. Schmalian,
Adv. Phys. {\bf 52}, 119 (2003).

\bibitem{Sandvik04} A. W. Sandvik, D. J. Scalapino, and N. E.
Bickers, Phys. Rev. B {\bf 69}, 094523 (2004).

\bibitem{Devereaux04} T. P. Devereaux, T. Cuk, Z.-X. Shen, and N.
Nagaosa, Phys. Rev. Lett. {\bf 93}, 117004 (2004).


\bibitem{Cren00} T. Cren {\em et al.}, Phys. Rev. Lett. {\bf 84}, 147
  (2000); Europhys. Lett. {\bf 54}, 84 (2001).

\bibitem{Howard01} C. Howald, P. Fournier, and A. Kaptitulnik,
  Phys. Rev. B {\bf 64}, 100504 (2001).

\bibitem{Pan01} S. H. Pan {\em et al.}, Nature {\bf 413}, 282 (2001).

\bibitem{Lang02} K. M. Lang {\em et al.}, Nature {\bf 415}, 412
  (2002).

\bibitem{Fang04} A. C. Fang {\em et al.}, Phys. Rev. B {\bf 70}, 214514
  (2004).

\bibitem{McElroy05} K. McElroy {\em et al.}, Science {\bf 309}, 1048
  (2005).

\bibitem{Jinho1} J. H. Lee et al., unpublished.
J. Lee, K. McElroy, J. Slezak, S. Uchida, H. Eisaki, and J.C. Davis,
Bull. Am. Phys. Soc. 50(2005) 299. J.C. Davis, J. Lee, K. McElroy,
J. Slezak, H. Eisaki, and S. Uchida, Bull. Am. Phys. Soc.
50(2005)1223.


\bibitem{Nunner05} T. S. Nunner, B. M. Anderson, A. Melikyan, and P.
J. Hirschfeld, Phys. Rev. Lett. {\bf 95}, 177003   (2005).

\bibitem{Norman:95} M.R. Norman, M. Randeria, H. Ding and J.C.
Campuzano, Phys. Rev. {\bf B 52}, 615, (1995).

\end{thebibliography}
\end{document}